\documentclass[12pt]{article}
\usepackage{amsfonts}
\usepackage{bm}
\usepackage{graphicx}
\usepackage{multicol}

\begin{document}

\title{\emph{}
Effect of noncommutativity on the spectrum of free particle and harmonic oscillator in rotationally invariant noncommutative phase space
}
\maketitle

\centerline {Kh. P. Gnatenko \footnote{E-Mail address: khrystyna.gnatenko@gmail.com}, O. V. Shyiko \footnote{E-Mail address: shijkoolga@ukr.net}}
\medskip

\centerline {\small  $^{1,2}$ \it  Ivan Franko National University of Lviv,}
\centerline {\small \it Department for Theoretical Physics,12 Drahomanov St., Lviv, 79005, Ukraine}

\begin{abstract}
We consider rotationally invariant noncommutative algebra with tensors of noncommutativity constructed with the help of additional coordinates and momenta. The algebra is equivalent to well known noncommutative algebra of canonical type.  In the noncommutative phase space with rotational symmetry influence of noncommutativity on the spectrum of free particle and spectrum of harmonic oscillator is studied up to the second order in the parameters of noncommutativity. We find that because of momentum noncommutativity the spectrum of free particle is discrete and corresponds to the spectrum  of harmonic oscillator in the ordinary space (space with commutative coordinates and commutative momenta). We obtain the spectrum of the harmonic oscillator in the rotationally invariant noncommutative phase space and conclude that noncommutativity of coordinates affects on its mass.  The frequency of the oscillator is affected by the coordinate noncommutativity and the momentum noncommutativity. On the basis of the results, the eigenvalues of squared length  operator  are found and restrictions on the value of length in noncommutative phase space with rotational  symmetry are analyzed.

\end{abstract}

\section{Introduction}

Recently, because of development of String Theory and Quantum Gravity (see, for example, \cite{Witten,Doplicher})  studies of idea of noncommutativity of coordinates on the Planck scale have received much attention.

Canonical version of noncommutative phase space is characterized by the following commutation relations
  \begin{eqnarray}
[X_{i},X_{j}]=i\hbar\theta_{ij},\label{form101}\\{}
[X_{i},P_{j}]=i\hbar(\delta_{ij}+\gamma_{ij}),\label{form1001}\\{}
[P_{i},P_{j}]=i\hbar\eta_{ij}.\label{form10001}{}
\end{eqnarray}
where $\theta_{ij}$, $\eta_{ij}$, $\gamma_{ij}$ are elements of constant matrixes. Parameters $\gamma_{ij}$ are considered to be defined as $\gamma_{ij}=\sum_k \theta_{ik}\eta_{jk}/4$  \cite{Bertolami,Bertolami3}.

 Many different problems were considered in noncommutative space of canonical type. Among them are free particle \cite{Djemai}, harmonic oscillator \cite{Hatzinikitas,Kijanka,Jing,Smailagic,Smailagic1,Djemai,Giri,Geloun,Nath,GnatenkoJPS17}, hydrogen atom \cite{Ho,Djemai,Chaichian,Chaichian1,Bertolami,Chair,Stern,Zaim2,Adorno,Khodja,Alavi,Gnatenko6,GnatenkoKr,GnatenkoConf,Balachandran}, Landau problem  \cite{Gamboa1,Horvathy,Dayi,Daszkiewicz1}, particle in gravitational field \cite{Bertolami1,Bastos,GnatenkoMPLA16},  many-particle systems \cite{Djemai,Ho,Daszkiewicz,Gnatenko1,Gnatenko2,Daszkiewicz2,Gnatenko9,GnatenkoMPLA17},  quantum fields  \cite{Balachandran10,Balachandran11} and others.

It is important to mention that in the case of  $\theta_{ij}$, $\eta_{ij}$, $\gamma_{ij}$ being elements of constant matrixes the commutation relations (\ref{form101})-(\ref{form10001}) are not rotationally invariant. The rotational symmetry is not preserved in noncommutative space of canonical type \cite{Chaichian,Balachandran1}.

Different noncommutative algebras were proposed in order to preserve rotational symmetry in noncommutative space (see, for instance, \cite{M,G,Amorim,GnatenkoPLA17,Gnatenko6}). Much attention has been devoted to studies of position-dependent noncommutativity  (see, for instance, \cite{Lukierski,Lukierski2009,Borowiec,Borowiec1,Kupriyanov2009,Kupriyanov} and reference therein).

In our paper \cite{GnatenkoIJMPA} we studied the problem of rotational symmetry breaking in noncommutative phase space and proposed noncommutative algebra  which is rotationally invariant and equivalent to noncommutative algebra of canonical type. For this purpose we considered the idea to construct tensors of noncommutativity with the help of  additional coordinates and momenta.

In the present paper we find  and analyze spectrum of free particle and spectrum of harmonic oscillator in rotationally invariant noncommutative phase space up to the second order in the parameters of noncommutativity. On the basis of the result the eigenvalues of the squared length operators defined in coordinate space, momentum space and phase space are obtained and  restrictions on the value of length in the space are analyzed.

The paper is organized as follows. In Section 2 noncommutative algebra which is rotationally invariant and equivalent to noncommutative algebra of canonical type is presented. Influence of noncommutativity on the free particle spectrum is analyzed in Section 3. In Section 4 we find the energy levels of harmonic oscillator in the noncommutative phase space with rotational symmetry. Section 5 is devoted to analysis of restrictions on the length in rotationally invariant noncommutative phase space.

\section{Rotationally invariant noncommutative phase space of canonical type}

In order to preserve rotational symmetry in noncommutative phase space in  recent papers \cite{GnatenkoIJMPA,GnatenkoIJMPA2} we considered idea to generalize parameters of noncommutativity to tensors
  \begin{eqnarray}
\theta_{ij}=\frac{c_{\theta} l^2_{P}}{\hbar}\sum_k\varepsilon_{ijk}\tilde{a}_{k}, \label{form130}\\
\eta_{ij}=\frac{c_{\eta}\hbar}{l^2_{P}}\sum_k\varepsilon_{ijk}\tilde{p}^b_{k}.\label{for130}
\end{eqnarray}
where $c_{\theta}$, $c_{\eta}$  are dimensionless constants, $l_P$ is the Planck length, $\tilde{a}_i$, $\tilde{b}_i$  $\tilde{p}^a_i$, $\tilde{p}^b_i$ are additional dimensionless coordinates and momenta conjugate of them which are governed by a rotationally symmetric systems.
As a result,  we proposed rotationally invariant noncommutative algebra
\begin{eqnarray}
[X_{i},X_{j}]=ic_{\theta} l^2_{P} \sum_k\varepsilon_{ijk}\tilde{a}_{k},\label{for101}\\{}
[X_{i},P_{j}]=i\hbar\left(\delta_{ij}+\frac{c_{\theta}c_{\eta}}{4}({\bf \tilde{a}}\cdot{\bf \tilde{p}^{b}})\delta_{ij}-\frac{c_{\theta}c_{\eta}}{4}{\tilde a}_j{\tilde p}^{b}_i\right),\label{for1001}\\{}
[P_{i},P_{j}]= \frac{c_{\eta}\hbar^2}{l_P^2}\sum_k\varepsilon_{ijk}\tilde{p}^{b}_{k}.{}\label{for10001}
\end{eqnarray}

For reason of simplicity we considered additional coordinates and additional momenta to be governed by harmonic oscillators
 \begin{eqnarray}
 H^a_{osc}=\hbar\omega_{osc}\left(\frac{(\tilde{p}^{a})^{2}}{2}+\frac{\tilde{a}^{2}}{2}\right),\label{form104}\\
 H^b_{osc}=\hbar\omega_{osc}\left(\frac{(\tilde{p}^{b})^{2}}{2}+\frac{\tilde{b}^{2}}{2}\right),\label{for104}
 \end{eqnarray}
 with  $\sqrt{\hbar}/\sqrt{{m_{osc}\omega_{osc}}}=l_{P}$ and very large frequency $\omega_{osc}$. So, harmonic oscillators put into the ground states remains in them because of very large distance between the energy levels.

Coordinates $\tilde{a}_{i}$, $\tilde{b}_{i}$ and momenta $\tilde{p}^{a}_{i}$, $\tilde{p}^{b}_{i}$ satisfy the ordinary commutation relations $[\tilde{a}_{i},\tilde{a}_{j}]=[\tilde{b}_{i},\tilde{b}_{j}]=[\tilde{a}_{i},\tilde{b}_{j}]=[\tilde{p}^{a}_{i},\tilde{p}^{a}_{j}]=[\tilde{p}^{b}_{i},\tilde{p}^{b}_{j}]=[\tilde{p}^{a}_{i},\tilde{p}^{b}_{j}]=0,$ $[\tilde{a}_{i},\tilde{p}^{a}_{j}]=[\tilde{b}_{i},\tilde{p}^{b}_{j}]=i\delta_{ij},$
 and
$[\tilde{a}_{i},\tilde{p}^{b}_{j}]=[\tilde{b}_{i},\tilde{p}^{a}_{j}]=0.$ Also, the coordinates and momenta commute with $X_i$, $P_i$, we have $[\tilde{a}_{i},X_{j}]=[\tilde{a}_{i},P_{j}]=[\tilde{p}^{b}_{i},X_{j}]=[\tilde{p}^{b}_{i},P_{j}]=0$.
As a result we have $[\theta_{ij}, X_k]=[\theta_{ij}, P_k]=[\eta_{ij}, X_k]=[\eta_{ij}, P_k]=[\gamma_{ij}, X_k]=[\gamma_{ij}, P_k]=0$.  So,  coordinates $X_{i}$, momenta $P_{i}$ and  $\theta_{ij}$, $\eta_{ij}$ satisfy the same commutation relations as in the case of the canonical version of noncommutative phase space  (\ref{for101})-(\ref{for10001}). Therefore, noncommutative algebra (\ref{for101})-(\ref{for10001}) is equivalent to noncommutative algebra of canonical type (\ref{form101})-(\ref{form10001}).

After rotation the coordinates and momenta reads $X_{i}^{\prime}=U(\varphi)X_{i}U^{+}(\varphi)$, $P_{i}^{\prime}=U(\varphi)P_{i}U^{+}(\varphi)$ $a_{i}^{\prime}=U(\varphi)a_{i}U^{+}(\varphi)$,  $p^{b\prime}_{i}=U(\varphi)p^b_{i}U^{+}(\varphi)$. Here $U(\varphi)=\exp(i\varphi({\bf n}\cdot{\bf L^t})/\hbar)$ and ${\bf L^t}$ is the total angular momentum defined as ${\bf L^t}=[{\bf x}\times{\bf p}]+\hbar[{\bf \tilde{a}}\times{\bf {\tilde p}}^{a}]+\hbar[{\bf {\tilde b}}\times{\bf {\tilde p}}^{b}]$ \cite{GnatenkoIJMPA}.
The commutation relations (\ref{for101})-(\ref{for10001}) remain the same after rotation, namely we have
\begin{eqnarray}
[X^\prime_{i},X^\prime_{j}]=ic_{\theta} l^2_{P} \sum_k\varepsilon_{ijk}\tilde{a}^\prime_{k},\\{}
[X^\prime_{i},P^\prime_{j}]=i\hbar\left(\delta_{ij}+\frac{c_{\theta}c_{\eta}}{4}({\bf \tilde{a}^\prime}\cdot{\bf \tilde{p}^{b\prime}})\delta_{ij}-\frac{c_{\theta}c_{\eta}}{4}{\tilde a}^\prime_j{\tilde p}^{b\prime}_i\right),\\{}
[P^{\prime}_{i},P^{\prime}_{j}]= \frac{c_{\eta}\hbar^2}{l_P^2}\sum_k\varepsilon_{ijk}\tilde{p}^{b\prime}_{k}.{}
\end{eqnarray}
So, noncommutative algebra (\ref{for101})-(\ref{for10001}) is rotationally invariant, besides it is equivalent to noncommutative algebra of canonical type.

 \section{Energy levels of free particle in rotationally invariant noncommutative phase space}
Let us consider free particle of mass $m$ in rotationally invariant noncommutative phase space (\ref{for101})-(\ref{for10001}). The hamiltonian of the particle reads
\begin{eqnarray}
H_{p}=\sum_i \frac{P_i^2}{2m}\label{fp000}.
\end{eqnarray}
Momenta $P_i$ do not commute. They satisfy (\ref{for10001}).

Because of involving of additional coordinates and additional momenta $\tilde{a}_{i}$, $\tilde{b}_{i}$, $\tilde{p}^{a}_{i}$, $\tilde{p}^{b}_{i}$, in rotationally invariant noncommutative space we have to consider the total hamiltonian which is defined as
\begin{eqnarray}
H=H_{p}+H^a_{osc}+H^b_{osc},\label{total}
\end{eqnarray}
where  $H^a_{osc}$, $H^b_{osc}$ are given by (\ref{form104}), (\ref{for104}).
Let us introduce operator
\begin{eqnarray}
\Delta H=H_{p}-\langle H_{p}\rangle_{ab}.\label{dd}
 \end{eqnarray}
 where we use notation $\langle...\rangle_{ab}$ for averaging over the eigenstates of oscillators (\ref{form104}), (\ref{for104}) in the ground states, $\langle...\rangle_{ab}=\langle\psi^{a}_{0,0,0}\psi^{b}_{0,0,0}|...|\psi^{a}_{0,0,0}\psi^{b}_{0,0,0}\rangle$. Here $\psi^{a}_{0,0,0}$, $\psi^{b}_{0,0,0}$  are eigenstates of  $H^a_{osc}$, $H^b_{osc}$ which are well known.
 So, the Hamiltonian (\ref{fp000}) can be written as
 \begin{eqnarray}
H=H_0+\Delta H,\label{total}
\end{eqnarray}
where $H_0$ reads
 \begin{eqnarray}
 H_0=\langle H_{p}\rangle_{ab}+H^a_{osc}+H^b_{osc}.\label{h0}
\end{eqnarray}

In the paper \cite{GnatenkoIJMPA2} it was shown that up to the second order of the perturbation theory corrections to the energy levels of $H_0$ caused by the term $\Delta H$ vanish. So, up to the second order in $\Delta H$, one can consider $H_0$ given by (\ref{h0}).

Let us calculate $\langle H_{p}\rangle_{ab}$. For this purpose it is convenient to use representation of noncommutative coordinates and noncommutative momenta by coordinates and momenta  $x_i$, $p_i$  which satisfy the ordinary commutation relations
 \begin{eqnarray}
[x_i,x_{j}]=[p_{i},p_{j}]=0,\label{cm}\\{}
[x_{i},p_{j}]=i\hbar\delta_{ij}.\label{cm1}
\end{eqnarray}
The representation is as follows
\begin{eqnarray}
X_{i}=x_{i}-\sum_j\frac{1}{2}\theta_{ij} p_j=x_{i}+\frac{1}{2}[{\bm \theta}\times {\bf p}]_i,\label{rrepx0}\\
P_{i}=p_{i}+\sum_j\frac{1}{2}\eta_{ij} x_j=p_{i}-\frac{1}{2}[{\bm \eta}\times {\bf x}]_i,\label{rrepp0}
\end{eqnarray}
where ${\bf x}=(x_1,x_2,x_3)$, ${\bf p}=(p_1,p_2,p_3)$. The components of vectors ${\bm \theta}$, ${\bm \eta}$ are defined as
\begin{eqnarray}
\theta_i=\sum_{jk}\varepsilon_{ijk}\frac{\theta_{jk}}{2}=\frac{c_{\theta} l^2_{P}}{\hbar}\tilde{a}_{i},\label{tt}\\
\eta_i=\sum_{jk}\varepsilon_{ijk}\frac{\eta_{jk}}{2}=\frac{c_{\eta} \hbar}{l_P^2}\tilde{p}^b_{i},\label{ee}
\end{eqnarray}
here we take into account (\ref{form130}) and (\ref{for130}).

So, using (\ref{rrepp0}) we can  write
\begin{eqnarray}
H_{p}=\frac{p^{2}}{2m}-\frac{({\bm \eta}\cdot[{\bf x}\times{\bf p}])}{2m}+\frac{[{\bm \eta}\times {\bf x}]^{2}}{8m}.\label{f222l02}
\end{eqnarray}
Note that $\langle H_{p}\rangle_{ab}=\langle\psi^{b}_{0,0,0}|H_p|\psi^{b}_{0,0,0}\rangle$ because $H_p$ does not depend on the $a_i$, $p^a_i$.  To find $\langle\psi^{b}_{0,0,0}|H_p|\psi^{b}_{0,0,0}\rangle$  we use the following results
\begin{eqnarray}
\langle\psi^{b}_{0,0,0}| \tilde{p}^b_i|\psi^{b}_{0,0,0}\rangle=0,\label{133}\\
\langle\psi^{b}_{0,0,0}| \tilde{p}^b_i\tilde{p}^b_j|\psi^{b}_{0,0,0}\rangle=\frac{1}{2}\delta_{ij}. \label{13333}
\end{eqnarray}
So, taking into account (\ref{ee}), we have
\begin{eqnarray}
\langle\eta_i\rangle_{ab}=0,\\
\langle\eta^2\rangle=\langle\eta^2\rangle_{ab}=\frac{3(\hbar c_{\eta})^2}{2 l_P^4}.
 \end{eqnarray}
Therefore after averaging, the linear term in the parameter of momentum noncommutativity  vanishes and we obtain
\begin{eqnarray}
\langle H_{p}\rangle_{ab}=\frac{p^2}{2m}+\frac{\langle\eta^2\rangle x^2}{12m}\label{fham1},
\end{eqnarray}
Using  (\ref{fham1}), we can also write
\begin{eqnarray}
\Delta H=-\frac{({\bm \eta}\cdot[{\bf x}\times{\bf p}])}{2m}+\frac{[{\bm \eta}\times {\bf x}]^{2}}{8m}-\frac{\langle\eta^2\rangle x^2}{12m}.\label{fbd}
\end{eqnarray}
Note, that $\Delta H$ contains terms of the first and second orders in the parameter of momentum noncommutativity. So, on the basis of conclusion presented in \cite{GnatenkoIJMPA2}, up to the second order in the parameter of momentum noncommutativity we can study $\langle H_{p}\rangle_{ab}$. Coordinates and momenta $x_i$, $p_j$ in $\langle H_{p}\rangle_{ab}$ satisfy the ordinary commutation relations. So, hamiltonian (\ref{fham1}) can be considered as hamiltonian of harmonic oscillator with mass $m$ and frequency
\begin{eqnarray}
\omega=\sqrt{\frac{\langle\eta^2\rangle}{6m^2}}.\label{frq}
\end{eqnarray}
and the spectrum of (\ref{fham1}) reads
 \begin{eqnarray}
E_{n_1,n_2,n_3}=\sqrt{\frac{\hbar^2\langle\eta^2\rangle}{6m^2}}\left(n_1+n_2+n_3+\frac{3}{2}\right),\label{sp}
\end{eqnarray}
with quantum numbers $n_1=0,1,2...$, $n_2=0,1,2...$, $n_3=0,1,2...$. So, the energy levels of free particle in the rotationally invariant noncommutative phase space correspond to the energy levels of tree-dimensional harmonic oscillator with frequency which depends on the parameter of momentum noncommutativity as (\ref{sp}). Noncommutativity of momenta causes quantization of the energy of free particle.

\section{Spectrum of harmonic oscillator in rotationally invariant noncommutative phase space}
Let us study the spectrum of three-dimensional harmonic oscillator with mass $m$  and frequency $\omega$
\begin{eqnarray}
H_{osc}=\sum_i \frac{P_i^2}{2m}+\sum_i \frac{m\omega^2X_i^2}{2}\label{222l02e}.
\end{eqnarray}
Operators $X_i$, $P_i$ satisfy commutation relations (\ref{for101})-(\ref{for10001}). The total hamiltonian reads
\begin{eqnarray}
H=H_0+\Delta H,\label{total}\\
H_0=\langle H_{osc}\rangle_{ab}+H^a_{osc}+H^b_{osc},\label{hh0}\\
\Delta H=H_{osc}-\langle H_{osc}\rangle_{ab}.
\end{eqnarray}
Using representation (\ref{rrepx0})-(\ref{rrepp0}) we can write
\begin{eqnarray}
H_{osc}=\frac{p^{2}}{2m}+\frac{m\omega^2x^2}{2}-\frac{({\bm \eta}\cdot[{\bf x}\times{\bf p}])}{2m}-\frac{m\omega^2({\bm \theta}\cdot[{\bf x}\times{\bf p}])}{2}+\nonumber\\+\frac{[{\bm \eta}\times {\bf x}]^{2}}{8m}+\frac{m\omega^2[{\bm \theta}\times {\bf p}]^{2}}{8}\label{222l02},
\end{eqnarray}
Let us find  $\langle H_{osc}\rangle_{ab}$. Taking into account (\ref{tt}), (\ref{ee}), (\ref{133}), (\ref{13333}) and
\begin{eqnarray}
\langle\psi^{a}_{0,0,0}| \tilde{a}_i|\psi^{a}_{0,0,0}\rangle=0,\label{13}\\
\langle\psi^{a}_{0,0,0}| \tilde{a}_i\tilde{a}_j|\psi^{a}_{0,0,0}\rangle=\frac{1}{2}\delta_{ij},\label{1333}\\
\end{eqnarray}
we obtain
\begin{eqnarray}
\langle H_{osc}\rangle_{ab}=\left(\frac{1}{2m}+\frac{m\omega^2\langle\theta^2\rangle}{12}\right)p^2+\left(\frac{m\omega^2}{2}+\frac{\langle\eta^2\rangle}{12m}\right)x^2\label{ham1},\\
\end{eqnarray}
here
\begin{eqnarray}
\langle\eta^2\rangle=\langle\eta^2\rangle_{ab}=\frac{3(\hbar c_{\eta})^2}{2 l_P^4}.
 \end{eqnarray}
We can also write
\begin{eqnarray}
\Delta H=-\frac{({\bm \eta}\cdot[{\bf x}\times{\bf p}])}{2m}-\frac{m\omega^2({\bm \theta}\cdot[{\bf x}\times{\bf p}])}{2}+\frac{[{\bm \eta}\times {\bf x}]^{2}}{8m}+\frac{m\omega^2[{\bm \theta}\times {\bf x}]^{2}}{8}-\nonumber\\-\frac{m\omega^2\langle\theta^2\rangle}{12}p^2-\frac{\langle\eta^2\rangle}{12m}x^2.\label{bd}
\end{eqnarray}
In $\Delta H$  one has  terms of the first and second orders in the parameters of noncommutativity.
 So,  up to the second order in the parameters of noncommutativity spectrum of the harmonic oscillator reads
 \begin{eqnarray}
 E_{n_1,n_2,n_3}=\hbar\sqrt{\left(m\omega^2+\frac{\langle\eta^2\rangle}{6m}\right)\left(\frac{1}{m}+\frac{m\omega^2\langle\theta^2\rangle}{6}\right)}\left(n_1+n_2+n_3+\frac{3}{2}\right)\label{sph}
 \end{eqnarray}
where $n_1$, $n_2$, $n_3$ are quantum numbers, $n_1=0,1,2...$, $n_2=0,1,2...$, $n_3=0,1,2...$ . So, up to the second order in the parameters of noncommutativity the spectrum of harmonic oscillator in rotationally invariant noncommutative space corresponds to the spectrum of harmonic oscillator in the ordinary space with effective mass and effective frequency. The mass of the oscillator in rotationally invariant noncommutative phase space is affected by the noncommutativity of coordinates. From (\ref{ham1}), we have
 \begin{eqnarray}
m_{eff}=\frac{6m}{6+m^2\omega^2\langle\theta^2\rangle}.\label{meff}
 \end{eqnarray}
The effective frequency of the oscillator depends on the parameters of coordinate and momentum noncommutativity as follows
  \begin{eqnarray}
\omega_{eff}=\sqrt{\left(m\omega^2+\frac{\langle\eta^2\rangle}{6m}\right)\left(\frac{1}{m}+\frac{m\omega^2\langle\theta^2\rangle}{6}\right)}.\label{omegaeff}
 \end{eqnarray}
 In the limits $\langle\theta^2\rangle\rightarrow0$, $\langle\eta^2\rangle\rightarrow0$ we have $m_{eff}=m$, $\omega_{eff}=\omega$ and expression (\ref{sph}) corresponds to the spectrum of the harmonic oscillator in the ordinary space.

 In the next section we will use this result to find restrictions on the length in rotationally invariant noncommutative phase space.

\section{Length in rotationally invariant noncommutative phase space}
Let us consider the squared length operator defined in phase space as follows
\begin{eqnarray}
{\bf Q}^2=\alpha^2\sum_i P_i^2+\beta^2\sum_i X_i^2\label{222l02e},
\end{eqnarray}
where $\alpha$ and $\beta$ being constants and $X_i$, $P_i$ satisfying commutation relations (\ref{form101})-(\ref{form10001}). The ratio $\alpha/\beta$ has dimension $s/kg$.
Operator (\ref{222l02e}) corresponds to Hamiltonian of tree-dimensional harmonic oscillator with mass $m=1/{2\alpha^2}$ and frequency $\omega=2\alpha\beta$. So, on the basis of the results presented in the previous section up to the second order in the parameters of noncommutativity  we can write eigenvalues of the operator  ${\bf Q}^2$ as follows
 \begin{eqnarray}
 q^2_{n_1,n_2,n_3}=\hbar\sqrt{\left(2\beta^2+\frac{\alpha^2\langle\eta^2\rangle}{3}\right)\left(2\alpha^2+\frac{\beta^2\langle\theta^2\rangle}{3}\right)}\left(n_1+n_2+n_3+\frac{3}{2}\right),\label{sppp}
 \end{eqnarray}
$n_1=0,1,2...$, $n_2=0,1,2...$, $n_3=0,1,2...$. From (\ref{sppp}) we have the following expression for the minimal length
 \begin{eqnarray}
 q^2_{min}=\sqrt{q^2_{0,0,0}}=\sqrt{\hbar}\sqrt[4]{2\beta^2+\frac{\alpha^2\langle\eta^2\rangle}{3}}\sqrt[4]{2\alpha^2+\frac{\beta^2\langle\theta^2\rangle}{3}}\label{rpmin}
 \end{eqnarray}
 So, the minimal length in noncommutative phase space is determined by the parameters of coordinate and momentum noncommutativity.
For $\alpha=0$, $\beta=1$ one has the squared length operator defined in the coordinate space
  \begin{eqnarray}
  {\bf R}^2=\sum^3_{i=1} X_i^2.
 \end{eqnarray}
 Taking into account (\ref{sppp}), the eigenvalues of the operator read
 \begin{eqnarray}
 r^2_{n_1,n_2,n_3}=\sqrt{\frac{2\hbar^2\langle\theta^2\rangle}{3}}\left(n_1+n_2+n_3+\frac{3}{2}\right),\label{spr}
 \end{eqnarray}
here $n_1=0,1,2...$, $n_2=0,1,2...$, $n_3=0,1,2...$. So, because of coordinate noncommutativity the squared length defined in the coordinate space is quantized. From (\ref{spr}) the minimal length is defined by the values of parameters of coordinate noncommutativity and reads
\begin{eqnarray}
r_{min}=\sqrt{r^2_{0,0,0}}=\sqrt{\frac{3\hbar^2\langle\theta^2\rangle}{2}}.\label{rmin}
\end{eqnarray}
 The result (\ref{sppp}) can be also used to analyze length defined in the momentum space.  In the case of $\alpha=1$, $\beta=0$, using (\ref{sppp}), we have
  \begin{eqnarray}
  {\bf P}^2=\sum^3_{i=1} P_i^2.\\
p^2_{n_1,n_2,n_3}=\sqrt{\frac{2\hbar^2\langle\eta^2\rangle}{3}}\left(n_1+n_2+n_3+\frac{3}{2}\right),\label{spp}
\end{eqnarray}
$n_1=0,1,2...$, $n_2=0,1,2...$, $n_3=0,1,2...$. So, in rotationally invariant noncommutative phase space the minimal length in the momentum space (minimal momentum) is defined as
\begin{eqnarray}
p_{min}=\sqrt{p^2_{0,0,0}}=\sqrt[4]{\frac{3\hbar^2\langle\eta^2\rangle}{2}}.\label{pmin}
\end{eqnarray}
The value of the minimal length in determined by the values of parameters of momentum noncommutativity.

\section{Conclusions}
In the paper we have considered noncommutative algebra (\ref{for101})-(\ref{for10001}) which is rotationally invariant and equivalent to noncommutative algebra of canonical type. The rotationally invariant algebra was proposed in \cite{GnatenkoIJMPA}. The algebra is constructed with the help of generalization of parameters of noncommutativity to tensors (\ref{form130}), (\ref{for130}) defined with the help of additional coordinates and momenta.

In rotationally invariant noncommutative phase space free particle has been considered.  We have found spectrum of the particle up to the second order in the parameter of momentum noncommutativity. It has been shown that because of momentum noncommutativity free particle has discrete energy spectrum (\ref{sp}). The spectrum corresponds to the spectrum of harmonic oscillator with frequency which is determined by the value of the parameter of momentum noncommutativity (\ref{frq}).

The influence of noncommutativaty of coordinates and noncommutativity of momenta on the spectrum of harmonic oscillator has been studied. We have found that noncommutativity effects on the mass and frequency of the oscillator (\ref{meff}), (\ref{omegaeff}). The spectrum of harmonic oscillator in rotationally invariant noncommutative space corresponds to the harmonic oscillator spectrum  with effective frequency (\ref{sph}) in the ordinary space.

The results have been used to analyze restrictions on the value of minimal length in rotationally invariant noncommutative phase space. We have found eigenvalues of squared length operator in the cases of its definition in the coordinate space, momentum space and phase space (\ref{sppp}), (\ref{spr}), (\ref{spp}). On the basis of the results the minimal length in the coordinate space (\ref{rmin}), minimal length in the momentum space (minimal momentum) (\ref{pmin}), and minimal length in phase space (\ref{rpmin}) have been obtained.

\section*{Acknowledgments}
 The authors thank Prof. V. M. Tkachuk and Dr. Yu. S. Krynytskyi for their advices and great support during research
studies. This work was partly supported by the project $\Phi\Phi$-63Hp (No. 0117U007190) from the Ministry of Education and Science of Ukraine.

\end{document}